\documentclass[english]{article}
\usepackage[T1]{fontenc}
\usepackage[latin9]{inputenc}
\usepackage[letterpaper]{geometry}
\usepackage{color,times}
\usepackage{array}
\usepackage{amsmath}
\usepackage{graphicx}
\usepackage{float}
\usepackage{cite}
\usepackage[unicode=true,
 bookmarks=false,
 breaklinks=false,colorlinks=false]
 {hyperref}

\makeatletter

\providecommand{\tabularnewline}{\\}

\usepackage{array}
\usepackage{textcomp}
\usepackage{fix-cm}

\providecommand{\tabularnewline}{\\}

\usepackage{babel}
\usepackage{esint}
\usepackage{epstopdf}
\usepackage{epsfig}

\makeatother

\begin{document}

\title{\textbf{Teleportation of a qubit using entangled non-orthogonal states:
A comparative study}}

\author{\textbf{Mitali Sisodia\footnote{Email: \href{mailto:mitalisisodiyadc@gmail.com}{mitalisisodiyadc@gmail.com}}, Vikram Verma\footnote{Email: \href{mailto:vikramverma18@gmail.com}{vikramverma18@gmail.com}}, Kishore
Thapliyal\footnote{Email: \href{mailto:tkishore36@yahoo.com}{tkishore36@yahoo.com}}, and Anirban Pathak\footnote{Email: \href{mailto:anirban.pathak@gmail.com}{anirban.pathak@gmail.com}}}\\ 
Jaypee Institute of Information Technology, A-10, Sector-62,
Noida, UP-201307, India}

\maketitle

\begin{abstract}
The effect of non-orthogonality of an entangled non-orthogonal state based
quantum channel is investigated in detail in the context of the teleportation of
a qubit. Specifically, average fidelity, minimum fidelity and minimum
assured fidelity (MASFI) are obtained for teleportation of a single
qubit state using all the Bell type entangled non-orthogonal states
known as quasi Bell states. Using Horodecki criterion, it is shown that the teleportation scheme obtained by replacing the quantum channel
(Bell state) of the usual teleportation scheme by a quasi Bell state
is optimal. Further, the performance of various quasi Bell states
as teleportation channel is compared in an ideal situation (i.e., in the
absence of noise) and under different noise models (e.g., amplitude and phase damping channels).
It is observed that the best choice of the quasi Bell state depends on
the amount non-orthogonality, both in noisy and noiseless case. A specific quasi Bell state, which was found to be maximally entangled in the ideal conditions, is shown to be less efficient as a teleportation channel compared to  other quasi Bell states in particular cases when subjected to noisy channels. It has also been observed that usually the value of average fidelity falls with an increase in the number of qubits exposed to noisy channels (viz., Alice's, Bob's and to be teleported qubits), but the converse may be observed in some particular cases. 
\end{abstract}

\section{Introduction\label{sec:Introduntion}}

One of the most important resources for quantum information processing
is entanglement. Entangled states and many of its applications do not
have any classical counterpart, and thus, an entangled state essentially
reveals and exploits nonclassicality (quantumness) present in a physical
system. For example, entanglement is essential (may not be sufficient)
for dense coding \cite{desnsecoding}, teleportation \cite{teleportation} and
device independent quantum cryptography \cite{DI-QKD}; and
none of these schemes (i.e., dense coding, teleportation, device independent
quantum cryptography) has any classical analog. Among these nonclassical
schemes, teleportation deserves special attention, as a large number
of quantum communication schemes can be viewed as variants of teleportation.
For example, quantum information splitting (QIS) \cite{QIS1,QIS2},
hierarchical QIS (HQIS) \cite{HQIS1,HQIS2}, quantum secret sharing
(QSS) \cite{QSS}, quantum cryptography based on entanglement swapping
\cite{ent-swap-based} may be viewed as variants of teleportation.
Usually, standard entangled states, which are inseparable states of
orthogonal states, are used to implement these teleportation based
schemes. However, entangled non-orthogonal states do exist and they
may be used to implement some of these teleportation-based protocols
\cite{15-satyabrata}. Specifically, entangled coherent states \cite{1-barry,2-Mann,3-Van,4-Peres,5-Mann,6-Wang92,7-Hprakash,8-Mishra,9-HPrakash}
and Schrodinger cat states prepared using $SU(2)$ coherent states
\cite{10-Wang} are the typical examples of entangled non-orthogonal
states. Such a state was first introduced by Sanders in 1992 \cite{1-barry}.
Since then several investigations have been made on the properties
and applications of the entangled non-orthogonal states. The investigations
have yielded a hand full of interesting results. To be precise, in Ref. \cite{hari-prakash-ent-swaping},
Prakash et al., have provided an interesting scheme for entanglement
swapping for a pair of entangled coherent states; subsequently, they
investigated the effect of noise on the teleportation fidelity obtained
in their scheme \cite{hprakash-noise-ent-swapping}, and Dong et
al. \cite{Dong}, showed that this type of entanglement swapping schemes
can be used to construct a scheme for continuous variable quantum
key distribution (QKD); the work of Hirota et al. \cite{Hirota},
has established that entangled coherent states, which constitute one
of the most popular examples of the entangled non-orthogonal states,
are more robust against decoherence due to photon absorption in comparison
to the conventional bi-photon Bell states; another variant of entangled
non-orthogonal states known as squeezed quasi Bell state has recently
been found to be very useful for quantum phase estimation \cite{quasi-bell-squeezed};
in \cite{15-satyabrata}, Adhikari et al., have investigated the merits and demerits of an entangled non-orthogonal state-based teleportation scheme analogous to the standard teleportation
scheme. In brief, these interesting works have established that most
of the quantum computing and communication tasks that can be performed
using usual entangled states of orthogonal states can also be performed
using entangled non-orthogonal states. 

The concept of MASFI, which was claimed to corresponds to the least
value of possible fidelity for any given information, was introduced
by Prakash et al. \cite{7-Hprakash}. Subsequently, in a series of papers (\cite{MASFI1,MASFI2} and references therein),
they have reported MASFI for various protocols of quantum communication,
and specially for the imperfect teleportation. Here, it is important
to note that Adhikari et al. \cite{15-satyabrata}, tried to extend 
the domain of the standard teleportation protocol to the case of performing
teleportation using entangled non-orthogonal states. To be precise,
they studied teleportation of an unknown quantum state by using a
specific type of entangled non-orthogonal state as the quantum channel.
In their protocol, the input state to be teleported from Alice to
Bob was prepared by a third party Charlie, who used to send the prepared
state to Alice though a noiseless quantum channel. Clearly, the presence
of Charlie was not essential, and in the absence of Charlie, their scheme
can be viewed simply as a scheme for teleportation using non-orthogonal
state. Interestingly, they s
d that the amount of non-orthogonality
present in the quantum channel affects the average fidelity ($F_{ave}$)
of teleportation. However, their work was restricted to a specific
type of entangled non-orthogonal state (a specific type of quasi Bell
state), and neither the optimality of the scheme nor the effect of
noise on it was investigated by them. In fact, in their work no effort
had been made to perform a comparative study (in terms of different
measures of teleportation quality) among possible quasi Bell states that can be used as teleportation channel. Further, the works of
Prakash et al., \cite{7-Hprakash,MASFI1,MASFI2} and others (\cite{MFI1,MFI2} and references therein) have established that in
addition to $F_{ave},$ minimum assured fidelity (MASFI) and minimum
average fidelity (MAVFI), which we refer here as minimum fidelity
(MFI) can be used as measures of quality of teleportation.  

Keeping these points in mind, in the present paper, we have studied
the effect of the amount of non-orthogonality on $F_{ave}$, MFI, and
MASFI for teleportation of a qubit using different quasi Bell states,
which can be used as the quantum channel. We have compared the performance
of these quasi Bell states as teleportation channel an ideal situation
(i.e., in the absence of noise) and in the presence of various types of noise
(e.g., amplitude damping (AD) and phase damping (PD)). The relevance of the choice of these noise models has been well established in the past (\cite{vishal,Kishore-decoy,crypt-switch} and references therein). Further, using Hordecky et al.'s relation \cite{Horodecki} between optimal
fidelity ($F_{opt}$) and maximal singlet fraction ($f$), it is established
that the entangled non-orthogonal state based teleportation scheme
investigated in the present work, is optimal for all the cases studied here (i.e.,
for all the quasi Bell states). 

The remaining part of the paper is organized as follows. In Sec. \ref{sec:Entangled-Non-orthngooal-States},
we briefly describe the mathematical structure of the entangled non-orthogonal
states and how to quantify the amount of entanglement present in such
states using concurrence. In this section, we have restricted ourselves
to very short description as most of the expressions reported here
are well known. However, they are required for the sake of a self
sufficient description. The main results of the present paper are reported
in Sec. \ref{sec:Quantum-Teleportation-using}, where we provide expressions
for ${\rm MASFI}$, $F_{ave}$, and $f$ for all the four quasi-Bell
states and establish $F_{ave}=F_{opt}$ for all the quasi Bell states,
and deterministic perfect teleportation is possible with the help
of quasi-Bell states. In Sec. \ref{sec:Effect-of-noise}, effects of amplitude damping and phase damping noise on $F_{ave}$ is discussed for various alternative situations, and finally the paper is concluded in Sec. \ref{sec:Conclusion}.

\section{Entangled non-orthogonal states\label{sec:Entangled-Non-orthngooal-States}}

Basic mathematical structures of standard entangled states and entangled
non-orthogonal states have been provided in detail in several papers
(\cite{1-barry,book} and references therein). Schmidt decomposition of an arbitrary bipartite state is written as
\begin{equation}
|\Psi\rangle=\underset{i}{\sum}p_{i}|\alpha_{i}\rangle_{A}\otimes|\beta_{i}\rangle_{B},
\end{equation}
where $p_{i}$s are the real numbers such that $\underset{i}{\sum}p_{i}^{2}=1$. Further, $\{|\alpha_{i}\rangle_{A}\}$ $\left(\{|\beta_{i}\rangle_{B}\}\right)$ is the orthonormal basis of subsystem $A$ $\left(B\right)$ in Hilbert space $H_{A}$ $\left(H_{B}\right)$. The state $|\Psi\rangle$ is entangled if at least two of the $p_{i}$s are non-zero. Here, we may note that a standard bipartite entangled
state can be expressed as 
\begin{equation}
|\psi\rangle=\mu|\alpha\rangle_{A}\otimes|\beta\rangle_{B}+\nu|\gamma\rangle_{A}\otimes|\delta\rangle_{B},\label{eq:1}
\end{equation}
where $\mu$ and $\nu$ are two complex coefficients that ensure normalization by satisfying $|\mu|^{2}+|\nu|^{2}=1$ in case of orthogonal states; $|\alpha\rangle$  and $|\gamma\rangle$
are normalized states of the first system and $|\beta\rangle$  and $|\delta\rangle$
are normalized states of the second system, respectively. These states of the subsystems satisfy   $\langle\alpha|\gamma\rangle=0$ and
$\langle\beta|\delta\rangle=0$ for the conventional entangled states of orthogonal states and they satisfy  $\langle\alpha|\gamma\rangle\neq0$ and
$\langle\beta|\delta\rangle\neq0$ for the entangled nonorthogonal states. 
Thus, an entangled
state involving non-orthogonal states, which is expressed in the form
of Eq. (\ref{eq:1}), has the property that the overlaps $\langle\alpha|\gamma\rangle$ and
$\langle\beta|\delta\rangle$ are nonzero, and the normalization condition
would be 
\begin{equation}
|\mu|^{2}+|\nu|^{2}+\mu\nu^{*}\langle\gamma|\alpha\rangle\langle\delta|\beta\rangle+\mu^{*}\nu\langle\alpha|\gamma\rangle\langle\beta|\delta\rangle=1.\label{eq:2}
\end{equation}
Here and in what follows, for simplicity, we have omitted the subsystems mentioned in the subscript.

The two non-orthogonal states of a given system are assumed to be
linearly independent and span a two dimensional subspace of the Hilbert
space. We may choose an orthonormal basis $\left\{ |0\rangle,|1\rangle\right\}$ as
\begin{equation}
|0\rangle=|\alpha\rangle,|1\rangle=\frac{(|\gamma\rangle-p_{1}|\alpha\rangle)}{N_{1}}\label{eq:3a}
\end{equation}
 for System $A$, and similarly, $|0\rangle=|\delta\rangle,{ |1\ensuremath{\rangle}=\ensuremath{\frac{(|\beta\rangle-p_{2}|\delta\rangle)}{N_{2}}} }$
for System $B$, where $p_{1}=\langle\alpha|\gamma\rangle,$ $p_{2}=\langle\delta|\beta\rangle$,
and $N_{i}=\sqrt{1-|p_{i}|^{2}}:\,i\in\{1,2\}.$ Now, we can express
the non-orthogonal entangled state $|\psi\rangle$ described by Eq. (\ref{eq:1})
using the orthogonal basis $\left\{ |0\rangle,|1\rangle\right\} $
as follows

\begin{equation}
|\psi\rangle=a|00\rangle+b|01\rangle+c|10\rangle,\label{eq:5}
\end{equation}
with $a=(\mu p_{2}+\nu p_{1})N_{12},\,b=(\mu N_{2})N_{12},\,c=(\nu N_{1})N_{12},$
where the normalization constant $N_{12}$ is given by

\begin{equation}
N_{12}=[|\mu|^{2}+|\nu|^{2}+\mu\nu^{*}\langle\gamma|\alpha\rangle\langle\delta|\beta\rangle+\mu^{*}\nu\langle\alpha|\gamma\rangle\langle\beta|\delta\rangle]^{-\frac{1}{2}}.\label{eq:normalization}
\end{equation}
Eq. (\ref{eq:5}) shows that an arbitrary entangled non-orthogonal
state can be considered as a state of two logical qubits. Following
standard procedure, the concurrence ($C$) \cite{11-Hill,12-Wooters}
of the entangled state $|\psi\rangle$ can be obtained as \cite{2-Mann,6-Wang92,13-Fu}

\begin{equation}
C=2|bc|=\frac{2|\mu||\nu|\sqrt{(1-|\langle\alpha|\gamma\rangle|^{2})(1-|\langle\beta|\delta\rangle|^{2})}}{|\mu|^{2}+|\nu|^{2}+\mu\nu^{*}\langle\gamma|\alpha\rangle\langle\delta|\beta\rangle+\mu^{*}\nu\langle\alpha|\gamma\rangle\langle\beta|\delta\rangle}.\label{eq:7}
\end{equation}

For the entangled state $|\psi\rangle$ to be maximally entangled,
we must have \textit{$C=1$}. Fu et al. \cite{13-Fu} showed that
the state $|\psi\rangle$ is maximally entangled state if and only
if one of the following conditions is satisfied: (i) $|\mu|=|\nu|$
for the orthogonal case, and (ii) $\mu=\nu e^{i\theta}$ and $\langle\alpha|\gamma\rangle=-\langle\beta|\delta\rangle^{*}e^{i\theta}$ for
the non-orthogonal states, where $\theta$ is a real parameter. 

Before we investigate the teleportation capacity of the entangled
non-orthogonal states, we would like to note that if we choose $\mu=\nu$
in Eq. (\ref{eq:1}), then for the case of orthogonal basis, normalization
condition will ensure that $\mu=\nu=\frac{1}{\sqrt{2}}$, and the state
$|\psi\rangle$ will reduce to a standard Bell state $|\psi^{+}\rangle=\frac{1}{\sqrt{2}}\left(|01\rangle+|10\rangle\right)$,
and its analogous state under the same condition (i.e., for $\mu=\nu$)
in non-orthogonal basis would be $|\psi_{+}\rangle=N_{+}\left(|\alpha\rangle\otimes|\beta\rangle+|\beta\rangle\otimes|\alpha\rangle\right),$
where $N_{+}$ is the normalization constant. In analogy to $|\psi^{+}\rangle$
its analogous entangled non-orthogonal state is denoted as $|\psi_{+}\rangle$
and referred to as quasi Bell state \cite{Hirota}. Similarly, in
analogy with the other 3 Bell states $|\psi^{-}\rangle=\frac{1}{\sqrt{2}}\left(|01\rangle+|10\rangle\right),\,|\phi^{+}\rangle=\frac{1}{\sqrt{2}}\left(|00\rangle+|11\rangle\right),$
and $|\phi^{-}\rangle=\frac{1}{\sqrt{2}}\left(|00\rangle-|11\rangle\right),$
we can obtain entangled non-orthogonal states denoted by $|\psi_{-}\rangle,\,|\phi_{+}\rangle,$
and $|\phi_{-}\rangle$, respectively. In addition to these notations,
in what follows we also use $|\psi^{+}\rangle=|\psi_{1}\rangle,\,|\psi^{-}\rangle=|\psi_{2}\rangle,\,|\phi^{+}\rangle=|\psi_{3}\rangle,\,|\phi^{-}\rangle=|\psi_{4}\rangle.$
Four entangled non-orthogonal states $\left\{ |\psi_{\pm}\rangle,\,|\phi_{\pm}\rangle\right\} $,
which are used in this paper, are usually referred to as quasi Bell
states \cite{Hirota}. They are not essentially maximally entangled,
and they may be expressed in orthogonal basis (see last column of
Table \ref{tab:quasi-bell-states}). Notations used in the rest of
the paper, expansion of the quasi Bell states in  orthogonal basis,
etc., are summarized in Table \ref{tab:quasi-bell-states}, where
we can see that $|\psi_{2}\rangle=|\psi^{-}\rangle$ is equivalent
to $|\psi_{-}\rangle,$ and thus $|\psi_{-}\rangle$ is always maximally
entangled and can lead to perfect deterministic teleportation as can
be done using usual Bell states. So $|\psi_{-}\rangle$ is not a state
of interest in noiseless case. Keeping this in mind, in the next section, we mainly
concentrate on the properties related to the teleportation capacity
of the other 3 quasi Bell states. However, in Sec. \ref{sec:Effect-of-noise}, we would discuss the effect of noise on all 4 quasi Bell states.

\begin{table}
\begin{centering}
\begin{tabular}{|c|c|>{\centering}p{5.5cm}|>{\centering}p{6.8cm}|}
\hline 
S. No. & Bell state & Corresponding Quasi Bell state 

(i.e., Bell-like entangled non-orthogonal state having a mathematical
form analogous to the usual Bell state given in the 2nd column of
the same row) & State in orthogonal basis that is equivalent to the quasi-Bell state
mentioned in the 3rd column of the same row\tabularnewline
\hline 
1. & $|\psi_{1}\rangle=|\psi^{+}\rangle$ & $|\psi_{+}\rangle=N_{+}(|\alpha\rangle\otimes|\beta\rangle+|\beta\rangle\otimes|\alpha\rangle),$ & $|\psi_{+}\rangle=\eta|00\rangle+\epsilon|01\rangle+\epsilon|10\rangle$\tabularnewline
\hline 
2. & $|\psi_{2}\rangle=|\psi^{-}\rangle$ & $|\psi_{-}\rangle=N_{-}(|\alpha\rangle\otimes|\beta\rangle-|\beta\rangle\otimes|\alpha\rangle),$ & $|\psi_{-}\rangle=\frac{1}{\sqrt{2}}\left(|01\rangle-|10\rangle\right)$\tabularnewline
\hline 
3. & $|\psi_{3}\rangle=|\phi^{+}\rangle$ & $|\phi_{+}\rangle={\rm M}_{+}(|\alpha\rangle\otimes|\alpha\rangle+|\beta\rangle\otimes|\beta\rangle)$ & $|\phi_{+}\rangle=k_{+}|00\rangle+l_{+}|01\rangle+l_{+}|10\rangle+m_{+}|11\rangle$\tabularnewline
\hline 
4. & $|\psi_{4}\rangle=|\phi^{-}\rangle$ & $|\phi_{-}\rangle={\rm M}_{-}(|\alpha\rangle\otimes|\alpha\rangle-|\beta\rangle\otimes|\beta\rangle)$ & $|\phi_{-}\rangle=k_{-}|00\rangle-l_{-}|01\rangle-l_{-}|10\rangle-m_{-}|11\rangle$\tabularnewline
\hline 
\end{tabular}
\par\end{centering}

\protect\caption{\label{tab:quasi-bell-states}Bell states and their analogous quasi
Bell states. The table also shows how quasi Bell states can be expressed
in orthogonal basis and introduces the notation used in this paper.
Here, $|\phi^{\pm}\rangle=\frac{1}{\sqrt{2}}\left(|00\rangle\pm|11\rangle\right)\,\,|\psi^{\pm}\rangle=\frac{1}{\sqrt{2}}\left(|01\rangle{\rm {\rm \pm}}|10\rangle\right)$,
$\eta=\frac{2re^{i\theta}}{\sqrt{2(1+r^{2})}},\,\epsilon=\sqrt{\frac{1-r^{2}}{2(1+r^{2})}}$,
$k_{\pm}=\frac{1\pm r^{2}e^{2i\theta}}{\sqrt{2(1\pm r^{2}\cos2\theta)}},\,\,l_{\pm}=\frac{(\sqrt{1-r^{2}})re^{i\theta}}{\sqrt{2(1\pm r^{2}\cos2\theta)}},\,\,m_{\pm}=\frac{1-r^{2}}{\sqrt{2(1\pm r^{2}\cos2\theta)}}$,
$N_{\pm}=\left[2\left(1\pm|\langle\alpha|\beta\rangle|^{2}\right)\right]^{\frac{-1}{2}},$ and $M_{\pm}=\frac{1}{\sqrt{2(1\pm r^{2}\cos2\theta)}}$
represent the normalization constant.}
\end{table}

\section{Teleportation using entangled non-orthogonal state\label{sec:Quantum-Teleportation-using} }

Let us consider that an arbitrary single qubit quantum state 

\begin{equation}
|I\rangle=x|0\rangle+y|1\rangle:\,\,\,\,\,\,\,\,\,\,\,\,\,\,\,|x|^{2}+|y|^{2}=1,\label{eq:8}
\end{equation}
is to be teleported using the quasi Bell state

\begin{equation}
|\psi_{\pm}\rangle=N_{\pm}(|\alpha\rangle\otimes|\beta\rangle\pm|\beta\rangle\otimes|\alpha\rangle),\label{eq:9}
\end{equation}
where the normalization constant $N_{\pm}=\left[2\left(1\pm|\langle\alpha|\beta\rangle|^{2}\right)\right]^{\frac{-1}{2}}$.
These quasi Bell states may be viewed as particular cases of Eq. (\ref{eq:1})
with $|\delta\rangle=|\alpha\rangle,$ $|\gamma\rangle=|\beta\rangle$,
and $\mu=\pm\nu$. In general, $\langle\alpha|\beta\rangle$ is a
complex number, and consequently, we can write

\begin{equation}
\langle\alpha|\beta\rangle=re^{i\theta},\label{eq:10}
\end{equation}
where the real parameters \textit{$r$} and \textit{$\theta$}, respectively,
denote the modulus and argument of the complex number $\langle\alpha|\beta\rangle$
with $0\leq r\leq1$ and $0\leq\theta\leq2\pi$. As $r=0$ implies,
orthogonal basis, we may consider this parameter as the primary measure
of non-orthogonality. This is so because no value of $\theta$ will
lead to orthogonality condition. Further, for $r\neq0,$ we can consider
$\theta$ as a secondary measure of non-orthogonality. Now, using
Eq. (\ref{eq:10}), and the map between orthogonal and non-orthogonal
bases we may rewrite Eq. (\ref{eq:3a}) as 

\begin{equation}
|0\rangle=|\alpha\rangle\,{\rm and}\,|1\rangle=\frac{\left[|\beta\rangle-\langle\alpha|\beta\rangle\alpha\rangle\right]}{\sqrt{1-r^{2}}}.\label{eq:11.a}
\end{equation}
Thus, we have $|\alpha\rangle=|0\rangle\,{\rm and}\,|\beta\rangle=\langle\alpha|\beta\rangle|0\rangle+\sqrt{1-r^{2}}|1\rangle,$
and consequently, $|\psi_{+}\rangle$ can now be expressed as

\begin{equation}
|\psi_{+}\rangle=\eta |00\rangle+\epsilon|01\rangle+\epsilon|10\rangle,\label{eq:12}
\end{equation}
where $\eta=\frac{2re^{i\theta}}{\sqrt{2(1+r^{2})}}$ and $\epsilon=\sqrt{\frac{1-r^{2}}{2(1+r^{2})}}$.
This is already noted in Table 1, where we have also noted that if
we express $|\psi_{-}\rangle$ in $\{|0\rangle,|1\rangle\}$ basis,
we obtain the Bell state $|\psi^{-}\rangle=\frac{1}{\sqrt{2}}\left(|01\rangle-|10\rangle\right)$,
which is maximally entangled and naturally yields unit fidelity for
teleportation. It's not surprising to obtain maximally entangled non-orthogonal
states, as in \cite{6-Wang92} it has been already established that
there exists a large class of bipartite entangled non-orthogonal states
that are maximally entangled under certain conditions. 

Using Eq. (\ref{eq:7}), we found the concurrence of the symmetric
state $|\psi_{+}\rangle$ as 

\begin{equation}
C\left(|\psi_{+}\rangle\right)=\frac{1-|\langle\alpha|\beta\rangle|^{2}}{1+|\langle\alpha|\beta\rangle|^{2}}=\frac{1-r^{2}}{1+r^{2}}.\label{eq:13}
\end{equation}
Clearly, $|\psi_{+}\rangle$ is not maximally entangled unless $r=\left|\langle\alpha|\beta\rangle\right|=0,$
which implies orthogonality. Thus, all quasi Bell states of the form
$|\psi_{+}\rangle$ are non-maximally entangled. Now, if the state
$|\psi_{+}\rangle$ is used as quantum channel, then following Prakash
et al. \cite{16-H-prakash}, we may express the MASFI for teleportation
of single qubit state (\ref{eq:8}) as 

\begin{equation}
\begin{array}{lcl}
\left({\rm MASFI}\right)_{\psi_{+}} & = & \frac{2C\left(|\psi_{+}\rangle\right)}{1+C\left(|\psi_{+}\rangle\right)}=1-r^{2}.\end{array}\label{eq:14}
\end{equation}

Since the value of $r$ lies between 0 and 1, the $\left({\rm MASFI}\right)_{\psi_{+}}$
decreases continuously as $r$ increases. For orthogonal state $r$=
$0$, and thus, ${\rm MASFI}=1$. Thus, we may conclude that the quasi
Bell state $|\psi_{+}\rangle$ will never lead to deterministic perfect
teleportation. However, its Bell state counter part ($r=1$ case)
leads to deterministic perfect teleportation. Here, it would be apt
to note that for teleportation of a single qubit state using $|\psi_{+}\rangle$
as the quantum channel, average teleportation fidelity can be obtained
as \cite{15-satyabrata}

\begin{equation}
F_{ave,\psi_{+}}=\frac{3-r^{2}}{3(1+r^{2})}.\label{eq:15}
\end{equation}
This is obtained by computing teleportation fidelity $F^{tel}=\sum_{i=1}^{4}P_{i}\left|\langle I|\zeta_{i}\rangle\right|^{2},$
where $|I\rangle$ is the input state, and $P_{i}=Tr(\langle\Omega|M_{i}|\Omega\rangle)$
with $|\Omega\rangle=|I\rangle\otimes|\psi_{\rm{channel}}\rangle$,
and $M_{i}=|\psi_{i}\rangle\langle\psi_{i}|$ is a measurement operator
in Bell basis ($|\psi_{i}\rangle$s are defined in the second column
of Table \ref{tab:quasi-bell-states}), and $|\zeta_{i}\rangle$ is the teleported state corresponding to $i$th projective measurement in Bell basis. Interestingly, $F^{tel}$
is found to depend on the parameters of the state to be teleported
(cf. Eq. (11) of Ref. \cite{15-satyabrata}). Thus, if we use Bloch
representation and express the state to be teleported as $|I\rangle=x|0\rangle+y|1\rangle=\cos\frac{\theta^{\prime}}{2}|0\rangle+\exp(i\phi^{\prime})\sin\frac{\theta^{\prime}}{2}|1\rangle$,
then the teleportation fidelity $F^{tel}$ will be a function of state
parameters $\theta^{\prime}$ and $\phi^{\prime}$ (here $^{\prime}$ is
used to distinguish the state parameter $\theta^{\prime}$ from the non-orthogonality
parameter $\theta$). An average fidelity is obtained by taking average
over all possible states that can be teleported, i.e., by computing
$F_{ave}=\frac{1}{4\pi}\int_{\phi^{\prime}=0}^{2\pi}\int_{\theta^{\prime}=0}^{\pi}F^{tel}\left(\theta^{\prime},\phi^{\prime}\right)\sin(\theta^{\prime})d\theta^{\prime}d\phi^{\prime}$.
This definition of average fidelity is followed in \cite{15-satyabrata,vederal}
and in the present paper.  However, in the works of Prakash et al.
(\cite{7-Hprakash,MASFI1,MASFI2} and references therein), $\left|\langle I|\zeta_{i}\rangle\right|^{2}$ was considered
as fidelity and $F^{tel}$ as average fidelity. They minimized $F^{tel}$
over the parameters of the state to be teleported and referred to
the obtained fidelity as the MAVFI. As that notation is not consistent
with the definition of average fidelity used here. In what follows,
we will refer to the minimum value of $F^{tel}$ as MFI, but it would
be the same as MAVFI defined by Prakash et al. Further, we would like
to note that in \cite{15-satyabrata} and in the present paper, it
is assumed that a standard teleportation scheme is implemented by
replacing a Bell state by its partner quasi Bell state, and as a consequence
for a specific outcome of Bell measurement of Alice, Bob applies the
same Pauli operator for teleportation channel $|\psi_{x}\rangle$ or
$|\phi_{x}\rangle$ (which is a quasi Bell state) as he used to do
for the corresponding Bell state $|\psi^{x}\rangle$ or $|\phi^{x}\rangle,$
where $x\in\{+,-\}.$ However, the expression of MASFI used here (see
Eq. (\ref{eq:14})) and derived in \cite{16-H-prakash} are  obtained using an optimized
set of unitary (cf. discussion after Eq. (10) in Ref. \cite{16-H-prakash}) and are subjected to outcome of Bell measurement of
Alice, thus no conclusions should be made by comparing MASFI with
MFI or $F_{ave}$.

From Eqs. (\ref{eq:14}) and (\ref{eq:15}), we can see that for a
standard Bell state $|\psi^{+}\rangle$ (i.e., when $r=0$), ${\rm (MASFI})_{\psi_{+}}=F_{ave}=1$.
However, for $r$ = 1, ${\rm (MASFI})_{\psi_{+}}=0,$  and $F_{ave}=\frac{1}{3}$. 
Thus, we conclude that for a standard Bell state both ${\rm MASFI}$
and average teleportation fidelity have the same value. This is not
surprising, as for $r$= 0 the entangled state $|\psi_{+}\rangle$
becomes maximally entangled. However, for $r\neq0,$ this state is
non-maximally entangled, and interestingly, for $r$= 1, we obtain
${\rm MASFI=0}$, whereas $F_{ave}$ is nonzero. We have already noted
that no comparison of ${\rm MASFI}$ and $F_{ave}$ obtained as above
should be made as that may lead to confusing results. Here we give
an example, according to \cite{7-Hprakash,16-H-prakash}, MASFI is
the least possible value of the fidelity, but for certain values of
$r$, we can observe that ${\rm MASFI}>F_{ave}.$ For example, for
$r=0.5$, we obtain ${\rm MASFI}=0.75,$ whereas $F_{ave}=0.733.$
Clearly, minimum found in computation of ${\rm MASFI}$, and the average
found in the computation of $F_{ave}$ is not performed over the same data
set, specifically not using the same teleportation mechanism (same unitary operations at the receiver's end). 

Now we may check the optimality of the teleportation scheme by using
the criterion introduced by Horodecki et al. in Ref. \cite{Horodecki}.
According to this criterion optimal average fidelity that can be obtained
for a teleportation schme which uses a bipartite entangled quantum
state $\rho$ as the quantum channel is 
\begin{equation}
F_{opt}=\frac{2f+1}{3},\label{eq:Horodecki}
\end{equation}
where $f$ is the maximal singlet fraction defined as 
\begin{equation}
f=\underset{i}{\max}\langle\psi_{i}|\rho|\psi_{i}\rangle,\label{eq:singlet fraction}
\end{equation}
where $|\psi_{i}\rangle$: $i\in\{1,2,3,4\}$ is Bell state described
above and summarized in Table \ref{tab:quasi-bell-states}. As we
are interested in computing $f$ for quasi Bell states which are pure
states, we can write $f=\underset{i}{\max}\left|\langle\psi_{i}|\chi\rangle\right|^{2},$ where $|\chi\rangle$ is a quasi Bell state. A bit of calculation
yields that maximal singlet fraction for the quasi Bell state $|\psi_{+}\rangle$
is 
\begin{equation}
f_{\psi_{+}}=\frac{1-r^{2}}{1+r^{2}}.\label{eq:singlerfracpsiplus}
\end{equation}
Now using (\ref{eq:15}), (\ref{eq:Horodecki}) and (\ref{eq:singlerfracpsiplus}), we can easily observe
that 
\begin{equation}
F_{opt,\,\psi_{+}}=\frac{2\left(\frac{1-r^{2}}{1+r^{2}}\right)+1}{3}=\frac{3-r^{2}}{3(1+r^{2})}=F_{ave}.\label{eq:optimality}
\end{equation}
Thus, a quasi Bell state based teleportation scheme which is analogous
to the usual teleportation scheme, but uses a quasi Bell state $|\psi_{+}\rangle$
as the quantum channel is optimal. We can also minimize $F_{\psi_{+}}^{tel}(\theta^{\prime},\phi^{\prime})$
with respect to $\theta^{\prime}$ and $\phi^{\prime}$ to obtain 
\begin{equation}
{\rm{MFI}}_{\psi_{+}}=\frac{1-r^{2}}{1+r^{2}},\label{eq:mfi-1}
\end{equation}
which is incidentally equivalent to maximal singlet fraction in this
case. 

So far we have reported analytic expressions for some parameters (e.g.,
$F_{ave},$ ${\rm MASFI},$ and ${\rm MFI)}$ that can be used as measures
of the quality of a teleportation scheme realized using the teleportation
channel $|\psi_{+}\rangle$ and have shown that the teleportation
scheme obtained using $|\psi_{+}\rangle$ is optimal. Among these
analytic expressions, $F_{ave,\psi_{+}}$ was already reported in \cite{15-satyabrata}.
Now, to perform a comparative study, let us consider that the teleportation
is performed using one of the remaining two quasi Bell states of our interest (i.e.,
using $|\phi_{+}\rangle$ or $|\phi_{-}\rangle$ described in Table
\ref{tab:quasi-bell-states}) as quantum channel. In that case, we
would obtain the concurrence as

\begin{equation}
C\left(|\phi_{\pm}\rangle\right)=2|\pm k_{\pm}m_{\pm}-l_{\pm}^{2}|=\frac{1-r^{2}}{(1\pm r^{2}\cos2\theta)}.\label{eq:19}
\end{equation}
Clearly, in contrast to $C|\psi_{+}\rangle,$ which was only $r$
dependent, the concurrence $C\left(|\phi_{\pm}\rangle\right)$ depends
on both the parameters $r$ and $\theta$. From Eq. (\ref{eq:19})
it is clear that at $\theta=\frac{\pi}{2},\frac{3\pi}{2}$ $\left(\theta=0,\pi\right)$
quasi Bell state $|\phi_{+}\rangle$ $\left(|\phi_{-}\rangle\right)$
is maximally entangled, even though the states $|\alpha\rangle$ and
$|\beta\rangle$ are non-orthogonal as $r\neq0$. Thus, at these points,
states $|\phi_{\pm}\rangle$ are maximally entangled. If quantum state
$|\phi_{+}\rangle$ is used as quantum channel, then ${\rm MASFI}$
for teleportation of an arbitrary single qubit information state (\ref{eq:8})
would be

\begin{equation}
{\rm (MASFI})_{\phi_{+}}=\frac{2C(|\phi_{+}\rangle)}{1+C(|\phi_{+}\rangle)}=\frac{1-r^{2}}{1-r^{2}\sin^{2}\theta},\label{eq:20}
\end{equation}
and similarly, that for quasi Bell state $|\phi_{-}\rangle$ would
be 

\begin{equation}
{\rm (MASFI})_{\phi_{-}}=\frac{1-r^{2}}{1-r^{2}\cos^{2}\theta}.\label{eq:21}
\end{equation}
Thus, the expressions for ${\rm MASFI}$ are also found to depend
on both $r$ and $\theta$. Clearly, at $\theta=\frac{\pi}{2}$ and
$\frac{3\pi}{2},$ ${\rm (MASFI})_{\phi_{+}}=1$, and hence for these
particular choices of $\theta,$ entangled non-orthogonal state $|\phi_{+}\rangle$
leads to the deterministic perfect teleportation of single qubit information
state. Clearly, for these values of $\theta$, $C(|\phi_{+}\rangle)=1,$
indicating maximal entanglement. However, the entangled state is still
non-orthogonal as $r$ can take any of its allowed values. Similarly,
at $\theta=0$ and $\pi,$ $({\rm MASFI})_{\phi_{-}}=1$, and hence
the entangled state $|\phi_{-}\rangle$ of the non-orthogonal states
$|\alpha\rangle$ and $|\beta\rangle$ leads to deterministic perfect
teleportation in these conditions. Thus, deterministic perfect teleportation
is possible using quasi Bell states $|\psi_{-}\rangle$ or $|\phi_{\pm}\rangle$
as quantum channels for teleportation, but it is not possible with
$|\psi_{+}\rangle$ unless it reduces to its orthogonal state counter
part (i.e., $|\psi^{+}\rangle).$ We may now compute the average fidelity for $|\phi_{\pm}\rangle$, by using the procedure adopted above for $|\psi_{+}\rangle$  and obtain 
\begin{equation}
F_{ave,\phi_{\pm}}=\frac{3-2r^{2}+r^{4}\mp r^{2}(r^{2}-3)\cos2\theta}{3(1\pm r^{2}\cos2\theta)}.\label{eq:avefidelityphipm}
\end{equation}

\begin{figure}
\centering{}\includegraphics[scale=0.8]{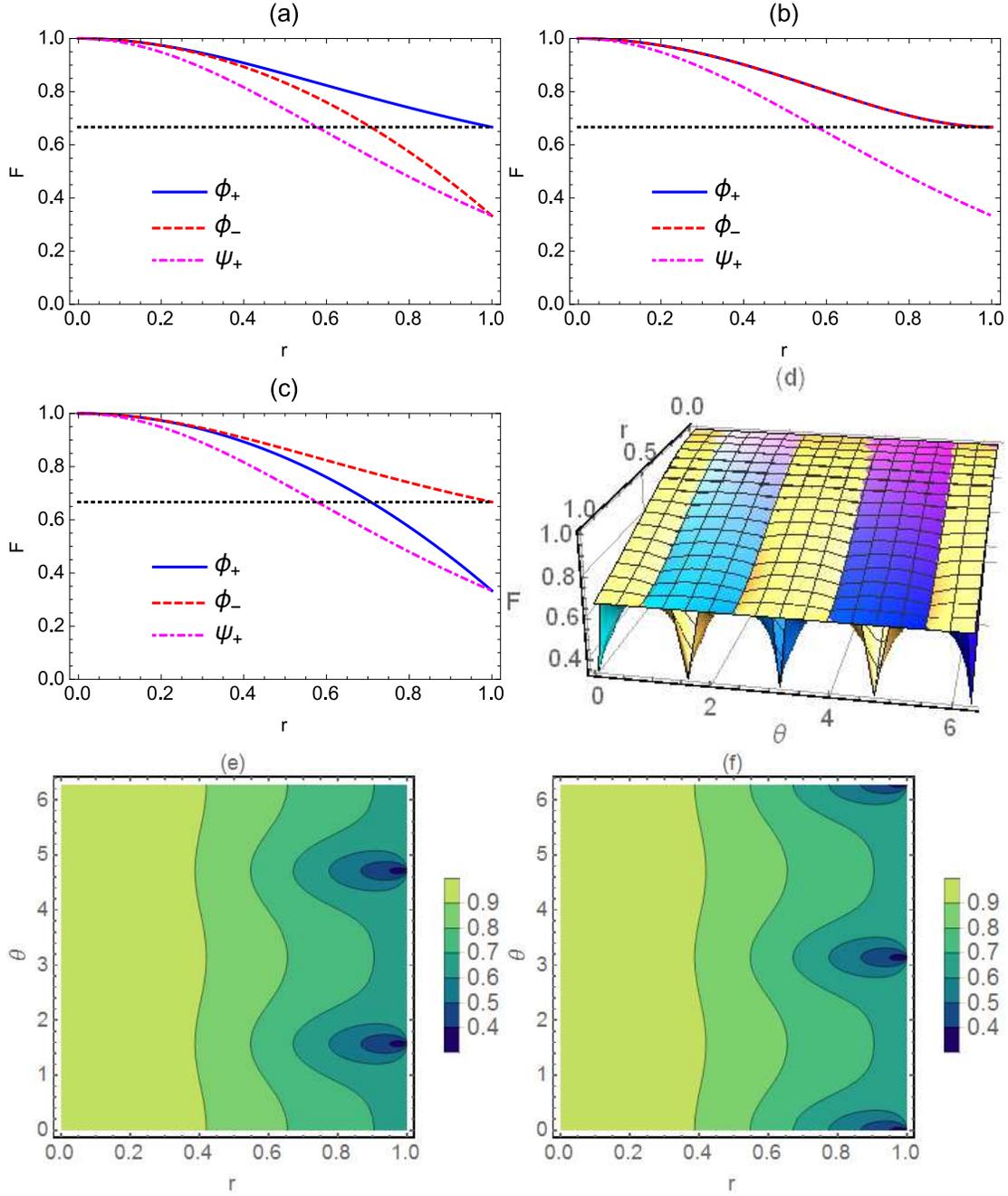}

\protect\caption{\label{fig:AvFid}(Color online) The dependence of the average fidelity on non-orthogonality parameters is established via 2 and 3 dimensional plots. Specifically, variation with $r$ for different values of $\theta=0,\frac{\pi}{4},$ and $\frac{\pi}{2}$ is shown in (a), (b), and (c), respectively. Similarly, (d) shows 3 dimensional variation for both $\phi_{+}$ and $\phi_{-}$ in light (yellow) and dark (blue) colored surface plots. The same fact can also be visualized using contour plots for both of these cases in (e) and (f). The horizontal dotted (black) line in (a), (b), and (c) corresponds to classical fidelity. Note that the quantity plotted in this and the following figures is $F_{ave}$, which is mentioned as $F$ in the $y$-axis.}
\end{figure}

Now, we would like to compare the average fidelity expressions obtained so far for various values of non-orthogonality parameters for all the quasi Bell states. The same is illustrated in  Fig. \ref{fig:AvFid}. Specifically, in Fig. \ref{fig:AvFid} (a)-(c) we have shown the dependence of average fidelity on secondary non-orthogonality parameter ($\theta$) using a set of plots for its variation with primary non-orthogonality parameter ($r$). This establishes that the primary non-orthogonality parameter has more control over the obtained average, fidelity but a secondary parameter is also prominent enough to change the choice of quasi Bell state to be preferred for specific value of primary non-orthogonality parameter. Thus, the amount of non-orthogonality plays a crucial role in deciding which quasi Bell state would provide highest average fidelity for a teleportation scheme implemented using quasi Bell state as the teleportation channel. Further, all these plots also establish that there is always a quasi Bell state apart from $|\psi_{-}\rangle$, which has average fidelity more than classically achievable fidelity $\left(\frac{2}{3}\right)$ for all values of $r$. We  may now further illustrate the dependence of the average fidelity on both non-orthogonality parameters via 3 D and contour plots shown in Fig. \ref{fig:AvFid} (d), (e), and (f). These plots establish that the average fidelity of $|\phi_{+}\rangle$ state increases for the values of $\theta$ for which $|\phi_{-}\rangle$ decreases, and vice-versa.

We can now establish the optimality of the teleportation scheme implemented using  $|\phi_{\pm}\rangle$ by computing average fidelity and maximal singlet fraction
for these channels. Specifically, computing the maximal singlet fraction
using the standard procedure described above, we have obtained 

\begin{equation}
f_{\phi_{\pm}}=\frac{2-2r^{2}+r^{4}\pm\cos2\theta(2r^{2}-r^{4})}{2(1\pm r^{2}\cos2\theta)}.\label{eq:singlet fractionphipm}
\end{equation}
Using Hordecky et al., criterion (\ref{eq:Horodecki}), and Eq. (\ref{eq:avefidelityphipm})-(\ref{eq:singlet fractionphipm}),
we can easily verify that $F_{ave,\phi_{\pm}}=F_{opt,\phi_{\pm}}$.
Thus, the teleportation scheme realized using any of the quasi Bell
state are optimal. However, they are not equally efficient for a specific choice of non-orthogonality parameter as we have
already seen in Fig. \ref{fig:AvFid}.  This motivates
us to further compare the performances of these quasi Bell states as a potential
quantum channel for teleportation. For the completeness of the comparative
investigation of the  teleportation efficiencies of different
quasi Bell states here we would also like to report MFI that can be achieved using different quasi Bell states. The
same can be computed as above, and the computation leads to following
analytic expressions of  MFI for $|\phi_{\pm}\rangle$: 
\begin{equation}
\begin{array}{lcl}
{\rm{MFI}}_{\phi_{+}} & = & \frac{1}{2}|k_{+}+m_{+}|^{2}=\frac{1-r^{2}(2-r^{2})\sin^{2}\theta}{1+ r^{2}\cos2\theta}\end{array}\label{eq:mavfi}
\end{equation}
and
\begin{equation}
\begin{array}{lcl}
{\rm{MFI}}_{\phi_{-}} & = & \frac{1}{2}|k_{-}+m_{-}|^{2}=\frac{1-r^{2}(2-r^{2})\cos^{2}\theta}{1- r^{2}\cos2\theta}\end{array}.\label{eq:mavfi}
\end{equation}

Interestingly, the comparative analysis performed with the expressions of MFI using their variation with various parameters led to quite similar behavior as observed for $F_{ave}$ in Fig. \ref{fig:AvFid}. Therefore, we are not reporting corresponding figures obtained for MFI.

\section{Effect of noise on average fidelity \label{sec:Effect-of-noise}}

In this section, we would like to analyze and compare the average fidelity obtained for each quasi Bell state over two well known Markovian channels, i.e., AD and PD channels. Specifically, in open quantum system formalism, a quantum state evolving under a noisy channel can be written in terms of Kraus operators as follows \cite{nielsen}
\begin{equation}
\rho\left(t\right)=\sum_{i}K_{i}\left(t\right)\rho\left(0\right)K_{i}^{\dagger}\left(t\right),\label{eq:noise-affected-density-matrix-1}
\end{equation}
where $K_{i}$s are the Kraus operators of a specific noise model. For example, in the case of AD channel the Kraus operators are \cite{nielsen}
\begin{equation}
K_{0}=|0\rangle\langle0|+\sqrt{1-\eta}|1\rangle\langle1|,\,\,\,\,\,\,\,\,\,\,\,K_{1}=\sqrt{\eta}|0\rangle\langle1|.\label{eq:Kraus-damping}
\end{equation}
Similarly, the Kraus operators for PD noise are \cite{nielsen}
\begin{equation}
\begin{array}{c}
K_{0}=|0\rangle\langle0|+\sqrt{1-\eta}|1\rangle\langle1|,\,\,\,\,\,\,\,\,\,\,\,K_{1}=\sqrt{\eta}|1\rangle\langle1|.\end{array}\label{eq:Kraus-dephasing}
\end{equation}
For both AD and PD noise, $\eta\,\left(0<\eta<1\right)$ is the decoherence rate, which determines the effect of the noisy channel on the quantum system.
To analyze the feasibility of quantum teleporation scheme using quasi Bell states and to compute the average fidelity we use Eqs. (\ref{eq:Kraus-damping}) and (\ref{eq:Kraus-dephasing}) in Eq. (\ref{eq:noise-affected-density-matrix-1}). Finally, to quantify the effect of noise we use a distance based measure "fidelity" between the quantum state evolved under a specific noisy channel and the quantum state Alice wish to teleport (say $\rho=|I\rangle\langle I|$). Mathematically, 

\begin{equation}
F_{k}=\frac{1}{4\pi}\int_{\phi^{\prime}=0}^{2\pi}\int_{\theta^{\prime}=0}^{\pi}\left(\sum_{i=1}^{4}P_{i}\langle I|\left\{\rho_{k}\left(\theta^{\prime},\phi^{\prime}\right)\right\}_{i}|I\rangle\right)\sin(\theta^{\prime})d\theta^{\prime}d\phi^{\prime},\label{eq:fidelity}
\end{equation}
which is the square of the conventional fidelity expression, and $\rho_{k}$ is the quantum state recovered at the Bob's port under the noisy channel $k\in\{AD,PD\}$. Further, details of the mathematical technique adopted here can be found in some of our recent works on secure \cite{vishal,Kishore-decoy,AQD} and insecure quantum communication \cite{crypt-switch,HJRSP}. 

We will start with the simplest case, where we assume that only Bob's part of the quantum channel is subjected to either AD or PD noise. The assumption is justified as the quasi Bell state used as quantum channel is prepared locally (here assumed to be prepared by Alice) and shared afterwards. During Alice to Bob transmission of an entnagled qubit, it may undergo decoherence, but the probability of decoherence is much less for the other qubits that don't travel through the channel (remain with Alice). Therefore, in comparison of the Bob's qubits, the Alice's qubits or the quantum state to be teleported $|I\rangle$, which remain at the sender's end, are hardly affected due to noise. The effect of AD noise under similar assumptions has been analyzed for three qubit GHZ and W states in the recent past \cite{our-QDs}.
The average fidelity for all four quasi Bell states, when Bob's qubit is subjected to AD channel while the qubits not traveling through the channel are assumed to be unaffected due to noise, is obtained as
\begin{equation}
\begin{array}{lcl}
F_{AD}^{|\phi_{+}\rangle} & = & \frac{-1}{2\left(3+3r^{2}\cos2\theta\right)}\left[-4+r^{2}\left(2+2\sqrt{1-\text{\ensuremath{\eta}}}-3\text{\ensuremath{\eta}}\right)-2\sqrt{1-\text{\ensuremath{\eta}}}\right.\\
& + & \left.2r^{4}(-1+\eta)+\text{\ensuremath{\eta}}+2r^{2}\left(-2-\sqrt{1-\text{\ensuremath{\eta}}}+r^{2}\sqrt{1-\text{\ensuremath{\eta}}}\right)\cos2\theta\right],\\
F_{AD}^{|\phi_{-}\rangle} & = & \frac{1}{-6+6r^{2}\cos2\theta}\left[-4+r^{2}\left(2+2\sqrt{1-\text{\ensuremath{\eta}}}-3\eta\right)-2\sqrt{1-\text{\ensuremath{\eta}}}\right.\\
& + & \left.2r^{4}(-1+\text{\ensuremath{\eta}})+\text{\ensuremath{\eta}}-2r^{2}\left(-2-\sqrt{1-\text{\ensuremath{\eta}}}+r^{2}\sqrt{1-\text{\ensuremath{\eta}}}\right)\cos2\theta\right],\\
F_{AD}^{|\psi_{+}\rangle} & = & \frac{4+2\sqrt{1-\text{\ensuremath{\eta}}}-\text{\ensuremath{\eta}}+r^{2}\left(-2\sqrt{1-\text{\ensuremath{\eta}}}+\eta\right)}{6\left(1+r^{2}\right)},\\
F_{AD}^{|\psi_{+}\rangle} & = & \frac{1}{6}\left[\left(4+2\sqrt{1-\eta}-\text{\ensuremath{\eta}}\right)\right].
\end{array}\label{eq:adsq}
\end{equation}

Here and in what follows, the subscript of fidelity $F$ corresponds to noise model and superscript represents the choice of quasi Bell state used as teleportation channel. Similarly, all the average fidelity expressions when Bob's qubit is subjected to PD noise can be obtained as

\begin{equation}
\begin{array}{lcl}
F_{PD}^{|\phi_{+}\rangle} & = & \frac{1}{3}\left[2+\sqrt{1-\eta}+r^{2}\left(-\sqrt{1-\eta}+\frac{-1+r^{2}}{1+r^{2}\cos2\theta}\right)\right],\\
F_{PD}^{|\phi_{-}\rangle} & = & \frac{1}{3}\left[2+\sqrt{1-\text{\ensuremath{\eta}}}-r^{2}\sqrt{1-\text{\ensuremath{\eta}}}+\frac{r^{2}-r^{4}}{-1+r^{2}\cos2\theta}\right],\\
F_{PD}^{|\psi_{+}\rangle} & = & \frac{2+\sqrt{1-\eta}-r^{2}\sqrt{1-\eta}}{3+3r^{2}},\\
F_{PD}^{|\psi_{-}\rangle} & = & \frac{1}{3}\left[2+\sqrt{1-\text{\ensuremath{\eta}}}\right].
\end{array}\label{eq:pdsq}
\end{equation}

It is easy to observe that for $\eta=0$ (i.e., in the absence of noise) the average fidelity expressions listed in Eqs. (\ref{eq:adsq}) and (\ref{eq:pdsq})  reduce to the average fidelity expressions corresponding to each quasi Bell state reported in Sec. \ref{sec:Quantum-Teleportation-using}. This is expected and can also be used to check the accuracy of our calculation.

It would be interesting to observe the change in fidelity when we consider the effect of noise on Alice's qubit as well. Though, it remains at Alice's port until she performs measurement on it in suitable basis, but in a realistic situation Alice's qubit may also interact with its surroundings in the meantime. Further, it can be assumed that the state intended to be teleported is prepared and teleported immediately. Therefore, it is hardly affected due to noisy environment. Here, without loss of generality, we assume that the decoherence rate for both the qubits is same. Using the same mathematical formalism adopted beforehand, we have obtained the average fidelity expressions for all the quasi Bell states when both the qubits in the quantum channel are affected by AD noise with the same decoherence rate. The expressions are

\begin{equation}
\begin{array}{lcl}
F_{AD}^{|\phi_{+}\rangle} & = & \frac{1}{3+3r^{2}\cos2\theta}\left[3-2r^{2}(-1+\text{\ensuremath{\eta}})^{2}+r^{4}(-1+\text{\ensuremath{\eta}})^{2}-2\eta\right.\\
& + & \left.\text{\ensuremath{\eta}}^{2}+r^{2}\left(3+r^{2}(-1+\text{\ensuremath{\eta}1})-\text{\ensuremath{\eta}}\right)\cos2\theta\right],\\
F_{AD}^{|\phi_{-}\rangle} & = & -\frac{1}{-3+3r^{2}\cos2\theta}\left[3-2r^{2}(-1+\text{\ensuremath{\eta}})^{2}+r^{4}(-1+\eta)^{2}+(-2+\text{\ensuremath{\eta}})\text{\ensuremath{\eta}}\right.\\
& + & \left.r^{2}\left(-3-r^{2}(-1+\text{\ensuremath{\eta}})+\text{\ensuremath{\eta}}\right)\cos2\theta\right],\\
F_{AD}^{|\psi_{+}\rangle} & = & \frac{3-2\eta+r^{2}(-1+2\eta)}{3\left(1+r^{2}\right)},\\
F_{AD}^{|\psi_{-}\rangle} & = & 1-\frac{2\text{\ensuremath{\eta}}}{3}.
\end{array}\label{eq:adbq}
\end{equation}

Similarly, the average fidelity expressions when both the qubits evolve under PD channel instead of AD channel are

\begin{equation}
\begin{array}{lcl}
F_{PD}^{|\phi_{+}\rangle} & = & \frac{1}{3}\left[3-\text{\ensuremath{\eta}}+r^{2}\left(-1+\text{\ensuremath{\eta}}+\frac{-1+r^{2}}{1+r^{2}\cos2\theta}\right)\right],\\
F_{PD}^{|\phi_{-}\rangle} & = & \frac{1}{3}\left[3+r^{2}(-1+\eta)-\eta+\frac{r^{2}-r^{4}}{-1+r^{2}\cos2\theta}\right],\\
F_{PD}^{|\psi_{+}\rangle} & = & \frac{3+r^{2}(-1+\text{\ensuremath{\eta}})-\text{\ensuremath{\eta}}}{3\left(1+r^{2}\right)},\\
F_{PD}^{|\psi_{-}\rangle} & = & 1-\frac{\eta}{3}.
\end{array}\label{eq:pdbq}
\end{equation}

\begin{figure}
\centering{}\includegraphics[scale=.65]{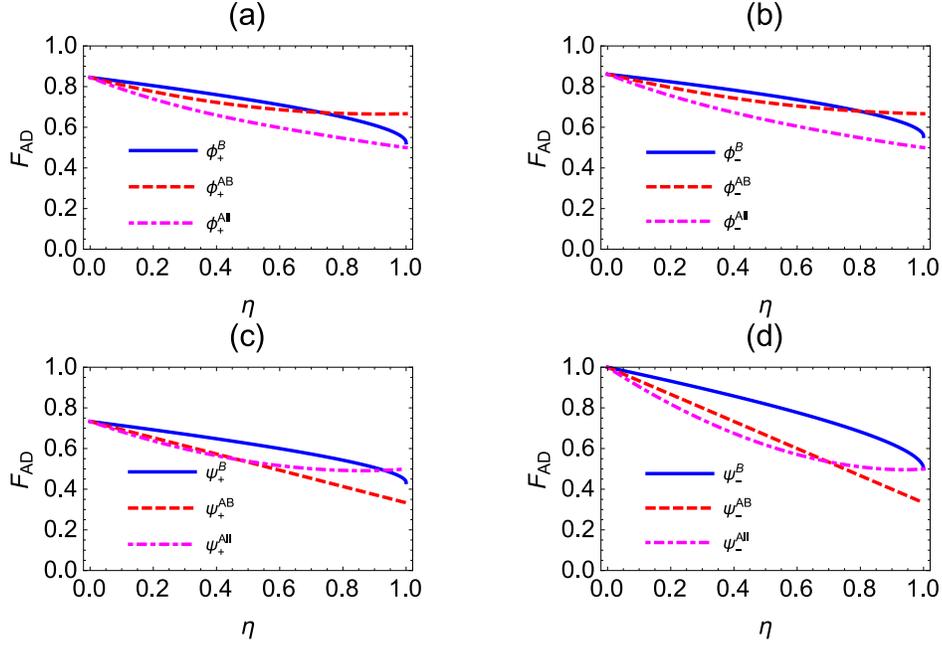}

\protect\caption{\label{fig:AvFid-AD}(Color online) The dependence of the average fidelity on the number of qubits exposed to AD channels is illustrated for $r=\frac{1}{2}$ and $\theta=\frac{\pi}{3}$. The choice of the initial Bell states in each case is mentioned in plot legends, where the superscript B, AB, and All corresponds to the cases when only Bob's, both Alice's and Bob's, and all three qubits were subjected to the noisy channel. The same notation is adopted in the following figures.}
\end{figure}

\begin{figure}
\centering{}\includegraphics[scale=.65]{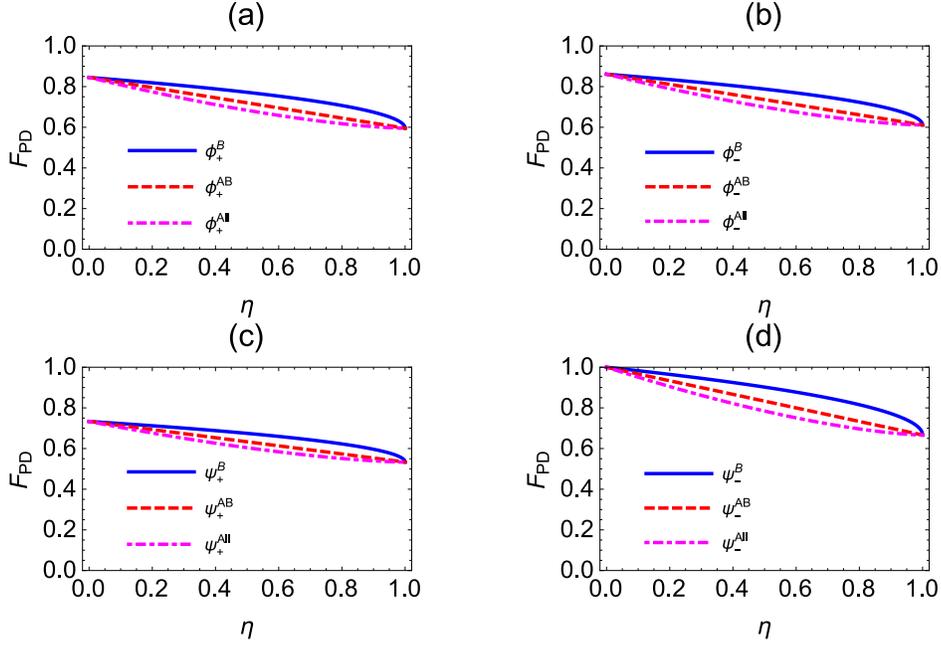}

\protect\caption{\label{fig:AvFid-PD}(Color online) Variation of the average fidelity for all the quasi Bell states is shown for all three cases with $r=\frac{1}{2}$ and $\theta=\frac{\pi}{3}$.}
\end{figure}

Finally, it is worth analyzing the effect of noisy channels on the feasibility of the teleportation scheme, when even the state to be teleported is also subjected to the same noisy channel. The requirement of this discussion can be established as it takes finite time before operations to teleport the quantum state are performed. Meanwhile, the qubit gets exposed to its vicinity and this interaction may lead to decoherence. Here, for simplicity, we have considered the same noise model for the state to be teleported as for the quantum channel. We have further assumed the same rate of decoherence for all the three qubits.
Under these specific conditions, when all the qubits evolve under AD channels, the average fidelity for each quasi Bell state turns out to be
\begin{equation}
\begin{array}{lcl}
F_{AD}^{|\phi_{+}\rangle} & = & \frac{-1}{2\left(3+3r^{2}\cos2\theta\right)}\left[-4+r^{2}\left(2+2\sqrt{1-\text{\text{\ensuremath{\eta}}}}-3\text{\text{\ensuremath{\eta}}}\right)-2\sqrt{1-\text{\text{\ensuremath{\eta}}}}+2r^{4}(-1+\text{\text{\ensuremath{\eta}}})\right.\\
& + & \left.\text{\text{\ensuremath{\eta}}}+2r^{2}\left(-2-\sqrt{1-\text{\ensuremath{\eta}}}+r^{2}\sqrt{1-\text{\ensuremath{\eta}}}\right)\cos2\theta\right],\\
F_{AD}^{|\phi_{-}\rangle} & = & \frac{1}{2\left(-3+3r^{2}\cos2\theta\right)}\left[-2\left(2+\sqrt{1-\text{\text{\ensuremath{\eta}}}}\right)+\text{\ensuremath{\eta}}\left(3+2\sqrt{1-\text{\text{\ensuremath{\eta}}}}+2(-2+\text{\text{\ensuremath{\eta}}})\text{\text{\ensuremath{\eta}}}\right)\right.\\
& - & 2r^{2}(-1+\text{\text{\ensuremath{\eta}}})\left(1+\sqrt{1-\text{\text{\ensuremath{\eta}}}}+\text{\text{\ensuremath{\eta}}}(-3+2\text{\ensuremath{\eta}})\right)+2r^{4}(-1+\text{\text{\ensuremath{\eta}}})^{3}\\
& + & \left.r^{2}\left(4+2\sqrt{1-\text{\text{\ensuremath{\eta}}}}+2\sqrt{1-\text{\text{\ensuremath{\eta}}}}\left(r^{2}(-1+\text{\ensuremath{\eta}})-\text{\text{\ensuremath{\eta}}}\right)-\text{\text{\ensuremath{\eta}}}\right)\cos2\theta\right],\\
F_{AD}^{|\psi_{+}\rangle} & = & \frac{1}{6\left(1+r^{2}\right)}\left[2\left(2+\sqrt{1-\text{\text{\ensuremath{\eta}}}}\right)+\text{\text{\ensuremath{\eta}}}\left(-3-2\sqrt{1-\text{\ensuremath{\eta}}}+2\text{\ensuremath{\eta}}\right)\right.\\
& + & \left.r^{2}\left(-2\sqrt{1-\text{\text{\ensuremath{\eta}}}}+\left(5+2\sqrt{1-\text{\ensuremath{\eta}}}-2\text{\text{\ensuremath{\eta}}}\right)\text{\text{\ensuremath{\eta}}}\right)\right],\\
F_{AD}^{|\psi_{-}\rangle} & = & \frac{1}{6}\left[4+2\sqrt{1-\text{\text{\ensuremath{\eta}}}}-3\text{\text{\ensuremath{\eta}}}-2\sqrt{1-\text{\ensuremath{\eta}}}\text{\text{\ensuremath{\eta}}}+2\text{\text{\ensuremath{\eta}}}^{2}\right].
\end{array}\label{eq:adaq}
\end{equation}

Similarly, when all three qubits are subjected to PD noise with the same decoherence rate, the analytic expressions of the  average fidelity are obtained as

\begin{equation}
\begin{array}{lcl}
F_{PD}^{|\phi_{+}\rangle} & = & \frac{1}{\left(3+3r^{2}\cos2\theta\right)}\left[2+r^{4}+\sqrt{1-\text{\ensuremath{\eta}}}-\sqrt{1-\text{\ensuremath{\eta}}}\text{\ensuremath{\eta}}+r^{2}\left(-1-\sqrt{1-\text{\ensuremath{\eta}}}+\sqrt{1-\text{\ensuremath{\eta}}}\eta\right)\right.\\
& + & \left.r^{2}\left(2+\sqrt{1-\text{\ensuremath{\eta}}}-r^{2}(1-\text{\ensuremath{\eta}})^{3/2}-\sqrt{1-\text{\ensuremath{\eta}}}\text{\ensuremath{\eta}}\right)\cos2\theta\right],\\
F_{PD}^{|\phi_{-}\rangle} & = & \frac{1}{3}\left[2+\sqrt{1-\text{\ensuremath{\eta}}}-r^{2}\sqrt{1-\text{\ensuremath{\eta}}}-\sqrt{1-\text{\ensuremath{\eta}}}\text{\ensuremath{\eta}}+r^{2}\sqrt{1-\text{\ensuremath{\eta}}}\text{\text{\ensuremath{\eta}}}+\frac{r^{2}-r^{4}}{-1+r^{2}\cos2\theta}\right],\\
F_{PD}^{|\psi_{+}\rangle} & = & \frac{2+\sqrt{1-\text{\text{\ensuremath{\eta}}}}+\sqrt{1-\text{\text{\ensuremath{\eta}}}}\left(r^{2}(-1+\text{\ensuremath{\eta}})-\text{\text{\ensuremath{\eta}}}\right)}{3\left(1+r^{2}\right)},\\
F_{PD}^{|\psi_{-}\rangle} & = & \frac{2+(1-\text{\text{\ensuremath{\eta}}})^{3/2}}{3}.
\end{array}\label{eq:pdaq}
\end{equation}

It is interesting to note that in the ideal conditions $|\psi_{-}\rangle$ is the unanimous choice of quasi Bell state to accomplish the teleportation with highest possible fidelity. However, from the expressions of fidelity obtained in Eqs. (\ref{eq:adsq})-(\ref{eq:pdaq}), it appears that it may not be the case in the presence of noise. For further analysis, it would be appropriate to observe the variation of all the fidelity expressions with various parameters. In what follows, we perform this analysis.

Fig. \ref{fig:AvFid-AD}, illustrates the dependence of the average fidelity on the number of qubits exposed to AD channel for each quasi Bell state using Eqs. (\ref{eq:adsq}), (\ref{eq:adbq}), and (\ref{eq:pdaq}). Unlike the remaining quasi Bell states, the average fidelity for $|\psi_{-}\rangle$ state starts from 1 at $\eta=0$. Until a moderate value (a particular value that depends on the choice of quasi Bell state) of decoherence rate is reached, the decay in average fidelity completely depends on the number of qubits interacting with their surroundings. However, at the higher decoherence rate, this particular nature was absent. Further, Fig. \ref{fig:AvFid-AD} (a) and (b) show that best results compared to remaining two cases can be obtained for the initial state $|\phi_{\pm}\rangle$, while both the channel qubits are evolving under AD noise; whereas the same case turns out to provide the worst results in case of $|\psi_{\pm}\rangle$. A similar study performed over PD channels instead of AD channels reveals that the decay in average fidelity solely depends on the number of qubits evolving over noisy channels (cf. Fig. \ref{fig:AvFid-PD}).

\begin{figure}

\centering{}\includegraphics[scale=0.85]{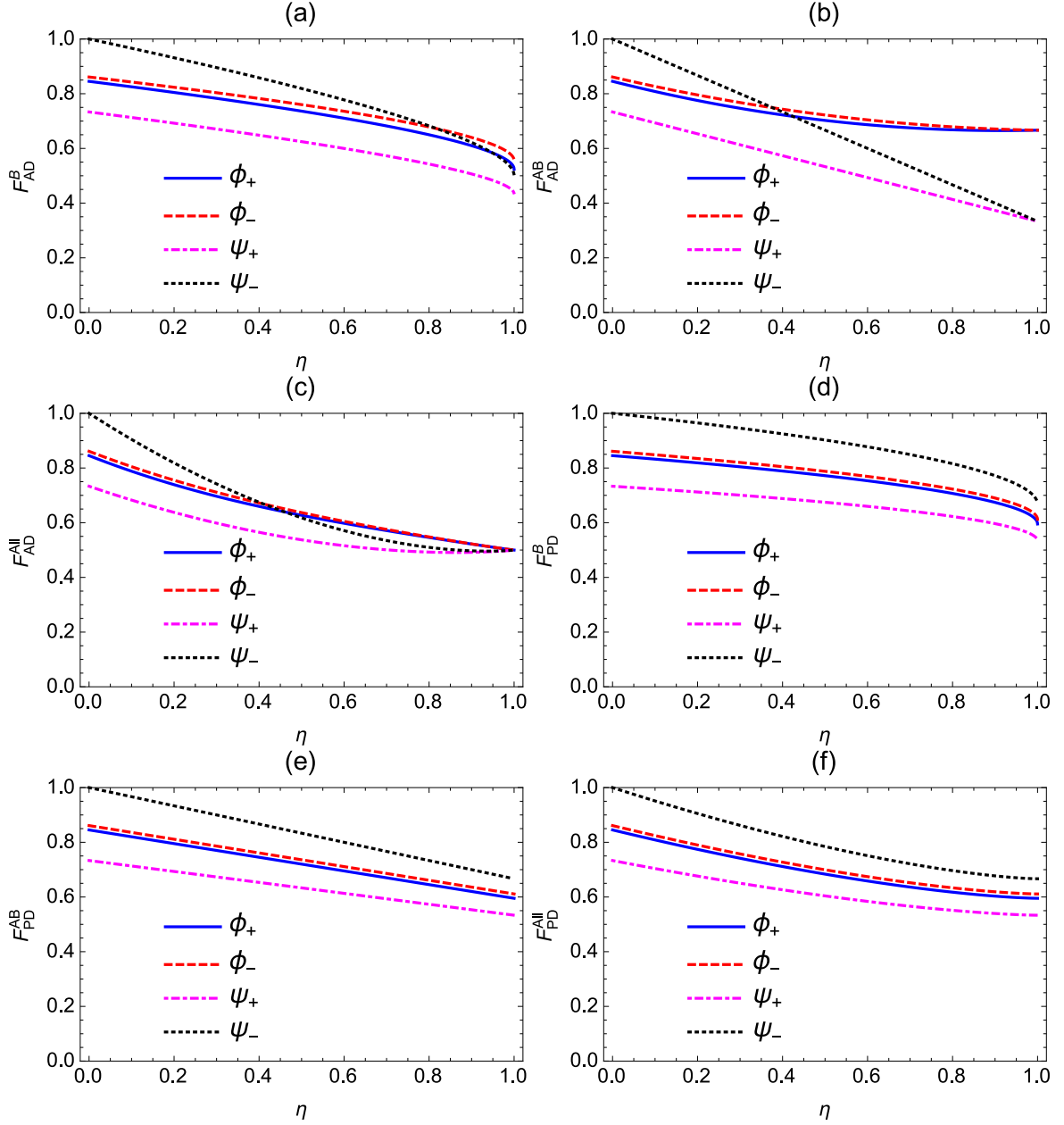}\protect\caption{\label{fig:AvFid-AD-PD}(Color online) The variation of average fidelity for all possible cases with each quasi Bell state as quantum channel is compared for AD ((a)-(c)) and PD ((d)-(f)) channels with $r=\frac{1}{2}$ and $\theta=\frac{\pi}{3}$. The case discussed has been mentioned as superscript of $F$ in the axes label on $y$-axis.}

\end{figure}

Finally, it is also worth to compare the average fidelity obtained for different quasi Bell states when subjected to noisy environment under similar condition. This would reveal the suitable choice of initial state to be used as a quantum channel for performing teleportation. In Fig. \ref{fig:AvFid-AD-PD} (a), the variation of average fidelity for all the quasi Bell states is demonstrated, while only Bob's qubit is exposed to AD noise. It establishes that although in ideal case and small decoherence rate $|\psi_{-}\rangle$ state is the most suitable choice of quantum channel, which do not remain true at higher decoherence rate. While  all other quasi Bell states follow exactly the same nature for decay of average fidelity and $|\psi_{+}\rangle$ appears to be the worst choice of quantum channel. A quite similar nature can be observed for the remaining two cases over AD channels in Fig. \ref{fig:AvFid-AD-PD} (b) and (c). Specifically, $|\psi_{-}\rangle$ remains the most suitable choice below moderate decoherence rate, while $|\phi_{\pm}\rangle$ may be preferred for channels with high decoherence, and $|\psi_{+}\rangle$ is inevitably the worst choice.

A similar study carried out over PD channels and the obtained results are illustrated in Fig. \ref{fig:AvFid-AD-PD} (d) and (f). From these plots, it may be inferred that $|\psi_{-}\rangle$ undoubtedly remains the most suitable and $|\psi_{+}\rangle$ the worst choice of quantum channel. The investigation on the variation of the average fidelity with non-orthogonality parameters over noisy channels yields a similar nature as was observed in ideal conditions (cf. Fig. \ref{fig:AvFid}). Therefore, we have not discussed it here, but a similar study can be performed in analogy with the ideal scenario. 

In fact, if one wishes to quantify only the effect of noise on the performance of the teleportation scheme using a non-orthogonal state quantum channel, the inner product may be taken with the teleported state in the ideal condition instead of the state to be teleported. The mathematical procedure adopted here is quite general in nature and would be appropriate to study the effect of generalized amplitude damping \cite{vishal,our-QDs}, squeezed generalized amplitude damping \cite{vishal,our-QDs}, bit flip \cite{vishal,HJRSP}, phase flip \cite{vishal,HJRSP}, and depolarizing channel \cite{vishal,HJRSP}. This discussion can further be extended to a set of non-Markovian channels \cite{non-Mark}, which will be carried out in the future and reported elsewhere.

\section{{Conclusion\label{sec:Conclusion}}}

In the present study, it is established that all the quasi Bell states, which are entangled non-orthogonal states, may  be used for quantum teleportation of a single qubit state. However, their teleportation efficiencies are not the same, and the efficiency  depends on the nature of noise present in the quantum channel. Specifically,  we have considered  four  quasi Bell states as possible teleportation channels, and  computed average and minimum fidelity that can be obtained by replacing a Bell state quantum channel of the standard teleportation scheme by its non-orthogonal counterpart (i.e., corresponding quasi Bell state). The results obtained here can be easily reduced to that obtained using usual Bell state in the limits of vanishing non-orthogonality parameter $r$. Specifically, there are two real parameters $r$ and $\theta$, which are considered  here as the primary and secondary measures of non-orthogonality, and variation of average and minimum fidelity is studied with respect to these parameters. Thus, in brief, the performance of the standard teleportation scheme is investigated using $F_{ave}$ and MFI as quantitative measures of quality of the teleportation scheme by considering a quasi Bell state instead of Bell state as quantum channel. Consequently, during discussion related to $F_{ave}$ and MFI, it has been assumed that Bob performs a Pauli operation corresponding to each Bell state measurement outcome as he used to perform in the standard  teleportation scheme. Further, using Horodecky criterion based on maximal singlet fraction it has been shown that such a scheme for quasi Bell state based teleportation is optimal.

We have used another  quantitative measure for quality of teleportation performance, MASFI, which is computed using a compact formula given in Ref. \cite{16-H-prakash}, where an optimal unitary operation is applied by Bob. For a few specific cases, the calculated MASFI was found to be unity. In those cases,  concurrence  for entangled non-orthogonal states were also found to be unity, which implied maximal entanglement. However, for this set of maximally entangled non-orthogonal states, we did not observe unit average fidelity and minimum fidelity as the unitary operations performed by Bob (used for computation of $F_{ave}$ and MFI ) were not the same as was in computation of MASFI.

The comparative study performed here led to a set of interesting observations that are illustrated in Figs. (\ref{fig:AvFid})-(\ref{fig:AvFid-AD-PD}). From Fig. (\ref{fig:AvFid}), we can clearly observe that there is at least a quasi Bell state (in addition to maximally entangled $|\psi_{-}\rangle$) for which the average fidelity that can be obtained in an ideal condition remains more than classical fidelity of teleportation. However, this choice of suitable quasi bell state completely depends on the value of secondary non-orthogonality parameter $\theta$, i.e., $\theta$ decides  whether $|\phi_{+}\rangle$ or $|\phi_{-}\rangle$ is preferable.
The performance of the teleportation scheme using entangled non-orthogonal states  has also been analyzed over noisy channels (cf. Figs. (\ref{fig:AvFid-AD})-(\ref{fig:AvFid-AD-PD})).  This study yield various interesting results. The quasi Bell state $|\psi_{-}\rangle$, which was shown to be maximally entangled in an ideal situation, remains most preferred choice as quantum channel while subjected to PD noise as well. However, in the presence of damping effects due to interaction with ambient environment (i.e., in AD noise), the choice of the quasi Bell state is found to depend  on the non-orthogonality parameter and the number of qubits exposed to noisy environment. 
 We hope the present study will be useful for experimental realization of teleportation schemes beyond usual entangled orthogonal state regime, and will also provide a clear prescription for future research  on applications of entangled non-orthogonal states. 
\color{black}

\textbf{Acknowledgement}: VV and AP
thank Department of Science and Technology (DST), India for the support
provided through the project number EMR/2015/000393.

\end{document}